\documentclass[showpacs,twocolumn,amssymb,prb,aps]{revtex4}
\usepackage{graphicx,epsfig}

\newcommand{\etal}{\textit{ et. al.}}

\begin{document}
\title{Liquid n-hexane condensed in silica nanochannels:\\ A combined optical birefringence and vapor sorption isotherm study}
\author{Andriy~V.~Kityk}
\email{andriy.kityk@univie.ac.at}
\affiliation{Department of Electrical Engineering, Czestochowa University of Technology,
Aleja Armii Krajowej 17, 42-200 Czestochowa, Poland}
\author{Klaus~Knorr}
\email{knorr@mx.uni-saarland.de}
\author{Patrick~Huber}
\email{p.huber@physik.uni-saarland.de}
\affiliation{Faculty of Physics and Mechatronics Engineering, Saarland University,
D-66041 Saarbr\"ucken, Germany}

\date{\today}

\begin{abstract}
The optical birefringence of liquid n-hexane condensed in an array of parallel silica channels of 7~nm diameter and 400~$\mu$m length is studied as a function of filling of the channels via the vapor phase. By an analysis with the generalized Bruggeman effective medium equation we demonstrate that such measurements are insensitive to the detailed geometrical (positional) arrangement of the adsorbed liquid inside the channels. However, this technique is particularly suitable to search for any optical anisotropies and thus collective orientational order as a function of channel filling. Nevertheless, no hints for such anisotropies are found in liquid n-hexane. The n-hexane molecules in the silica nanochannels are totally orientationally disordered in all condensation regimes, in particular in the film growth as well as in the the capillary condensed regime. Thus, the peculiar molecular arrangement found upon freezing of liquid n-hexane in nanochannel-confinement, where the molecules are collectively aligned perpendicularly to the channels` long axes, does not originate in any pre-alignment effects in the nanoconfined liquid due to capillary nematization.
\end{abstract}

\pacs{68.03.Fg, 61.46.-w, 42.25.Lc}

\maketitle

\section{Introduction}

The structural and thermodynamical properties of liquids spatially restricted in at least one direction to a few nanometers or confined in semi-infinite manner at liquid surfaces and liquid/solid interfaces can differ markedly in comparison to the bulk state \cite{LitIntroExp, LitIntroTheo}. For example, surface or interfacial induced layering of liquids has been reported for many systems ranging from simple liquid metals, linear hydrocarbons and ionic liquids to ferrofluids \cite{HarvardLMs, Volkmann2002,vanderVeen1997, Mezger2008}.

Therefore, quite naturally the question arises to what extent the liquid state may be affected by the interaction with the channel walls, that is by the liquid/solid interfacial properties, when liquids are confined in channels or pores of a few nanometer in diameter. In fact, it has been established for a long time that liquids confined in nanopores can thermodynamically be separated in two components, a few monolayers in the proximity of the pore walls which strongly interact with the surrounding matrix and a second component in the pore center, which is established via capillary condensation \cite{Ev, Saam, Huber1999}. Aside from the influence of the confining walls on the translational order in the liquid, the question arises if spatial confinement can induce preferred orientational order in liquids comprising anisotropic building blocks, that is capillary nematization \cite{CapillaryNematization}. In fact, there has been produced substantial evidence for such effects for a variety of liquid crystals, where spatial confinement on the nanoscale can induce nematic or paranematic order \cite{Crawford1996, Kityk2008}. Analogous effects may also occur for liquids built out of molecules with less pronounced shape anisotropy, e.g. medium-length linear hydrocarbons \cite{Klein1995, Valiullin2006}.

This question may be considered as of particular interest for this class of molecules, since upon crystallization of melts of medium-length n-alkanes (and n-alcohols) a distinct collective orientational order has been observed in x-ray diffraction experiments \cite{Henschel2007}: The long axes of the molecules are all arranged perpendicularly to the channels' long axes, despite a sizeable channel wall roughness in the mesoporous silicon matrices employed in these studies. The peculiar texture was attributed to a nanoscale version of the mechanism underlying the Bridgman technique of single-crystal growth \cite{Bridgman1925}. It predicts an alignment of fast growing crystalline directions along the channels' long axis, which in the case of n-alkanes are directions perpendicular to the molecules' backbones. Consequently, it dictates a somewhat counter-intuitive molecular arrangement, where the alkane`s long axes are oriented perpendicularly to the long axes of the channels. The question, however, whether the peculiar texture in the solid state is solely dictated by the nano-scale Bridgman mechanism or whether pre-alignment effects in the nanoconfined liquid, as discussed above, may at least partially be responsible for the effectiveness of this mechanism, has so far remained unanswered.

\begin{figure}[htbp]
\epsfig{file=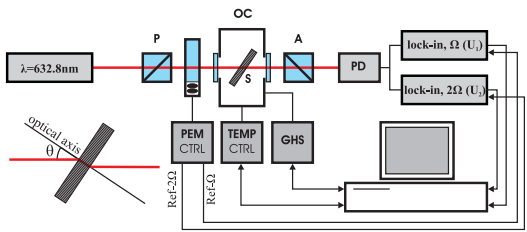, angle=0, width=0.8\columnwidth}\caption{(color online) Schematic of a high-resolution optical birefringence setup consisting of a He-Ne
laser, an optical polarizer (P), a temperature controlled optical cell (OC), the sample (S), an optical analyzer (A), a photoelastic modulator
(PEM-90), a photodiode as a detector (PD), and a ''lock-in'' detection and analyzing unit. The inset shows the orientation of the solid-state silica membrane with respect to the incident laser beam.
} \label{fig1}
\end{figure}

Therefore, the aim of this study is to scrutinize for possible confinement induced anisotropies in liquid n-hexane, a simple, slightly anisotropic molecule (molecular length $\sim$ 0.8~nm, molecular width $\sim$ 0.25~nm), where in the spatially nanochannel-confined crystalline state a preferential collective orientational order, as outlined above, has been observed. To this end we performed optical birefringence measurements as a function of the n-hexane filling fraction $f$ of an array of tubular silica channels of a few nanometer in diameter and 400~$\mu$m length upon ad- and desorption of the liquid via the vapor phase.

Unfortunately such measurements cannot be applied to conventional disordered mesoporous substrates. In Vycor glass and the xerogels termed ''controlled pore glasses'' the pores are oriented randomly. In template grown varieties of silica, e.g. SBA-15 \cite{Zhao1998, Hofmann1995}, the pores are oriented all parallel, but since these samples are powders, there is again no macroscopic alignment of the pores or channels, and thus any optical anisotropies which may originate from orientational molecular ordering are washed out. A macroscopic alignment of the tubular pores does exist however for electrochemically etched mesoporous Si. In mesoporous Si (100) sheets, the long axes of the tubular pores are perpendicular to the surface. Upon oxidation such substrates keep the macroscopic alignment and become insulators, transparent for visible light \cite{MesoSiOxy}. Thus oxidized porous silicon is an ideal substrate for the study of confinement induced optical and dielectric anisotropies in liquid condensates. As we shall demonstrate $f$-dependent birefringence data from this matrix can be analyzed in terms of simple effective medium expressions. In particular, we test different models with different geometric partitions and optical anisotropies in order to explore the molecular orientational order in the channels.

\section{Experimental}

Mesoporous Si \cite{MesoSi} was prepared by electrochemical etching of highly $p$-doped (100) wafer with a resistivity of 0.01~$\Omega \cdot$cm. The ultrasonically cleaned wafer was mounted in a Teflon cell and exposed to a solution of electrolyte solution of 48\% hydrofluoric acid and 98\% ethanol taken in volume fractions 3:7. An etching current of 12.5 mA/cm$^2$ was applied for 400~minutes. The resulting porous layer was released from the bulk Si underneath by increasing the etching current by a factor of 10. Subsequently the porous Si sheet was oxidized at 800 $^o$C in ambient atmosphere \cite{MesoSiOxy}. Standard analysis of a $N_{\rm 2}$ sorption isotherm recorded at 77~K yields a porosity $P$ of 48\% and a pore diameter $2R_{\rm 0}$ of 7~nm. The sheet is 397$\pm$5~$\mu$m thick, hence the aspect ratio channel length to channel diameter is almost 10$^5$. Electron micrographs of the initial porous silicon channels \cite{Gruener2008, Kumar2008} indicate sizeable 1.0$\pm0.5$~nm mean square deviations of their inner surfaces from an ideal cylindrical form. Similar sizeable values of the channel roughness have been inferred from electron micrographs and AFM pictures of membranes thermally oxidized under similar conditions as employed here \cite{MesoSiOxyRoughness}.

Because of the alignment of the channels perpendicular to the surface, the wafer is an effective medium that shows an uniaxial ''shape'' anisotropy, even in the case that the constituents have isotropic permittivities and refractive indices. This is analogous to the observations for mesoporous silicon\cite{Kunzner2005}, aligned carbon nanotube films \cite{Heer}, porous Si \cite{Gen,Kunz} or semiconductor nanowires \cite{Mus}. The resulting birefringence can be determined from the phase difference $R$ between two perpendicularly polarized light beams (polarimetry).

We have used a high-resolution optical polarization method for the accurate determination of the retardation $R$. The setup (see Fig.~\ref{fig1}) employs an optical photoelastic modulator (PEM-90, Hinds Instrument) in combination with a dual lock-in detection scheme to minimize the effects of uncontrolled light-intensity fluctuations \cite{Skarabot}.  The intensity of the modulated laser light ($\lambda$=632.8 nm) was detected by a photodiode and two lock-in amplifiers, which simultaneously determined the amplitudes of the first ($U_{\rm \Omega}$) and second ($U_{\rm 2\Omega}$) harmonics, respectively. The retardation of the sample $R=\arctan[(U_{\rm \Omega}J_{\rm2}(A_{\rm0}))/(U_{\rm 2\Omega}J_{\rm 1}(A_{\rm 0}))]$ (here $J_{\rm 1}(A_{\rm 0})$ and $J_{\rm 2}(A_{\rm 0})$ are the Bessel functions corresponding to the PEM retardation amplitude $A_{\rm 0}=0.383\lambda$) was measured for the light incident at an angle $\theta=38.7$ deg (see insert, Fig.~\ref{fig1}) with respect to the membrane surface. The sample have been kept in a closed optical chamber (OC), held at a constant temperature ($T=$281~K) and small volumetrically controlled portions of n-hexane vapor have been subsequently added (for adsorption) or removed (for desorption) from the chamber by means of an all-metal gas handling system (GHS) equipped with a capacitive membrane gauge of 100~mbar full scale. The equilibration time was two hours for each step. The pore condensate is specified by the reduced pressure $p=P/P_{\rm 0}$ and filling fraction $f=N/N_{\rm 0}$. $P_{\rm 0}$ is the saturated vapor pressure of bulk n-hexane at given temperature $T$, $N$ is the number of adsorbed molecules in the porous substrate and $N_{\rm 0}$ is the number of molecules necessary for complete filling of the pores.

\begin{figure}[htbp]
\epsfig{file=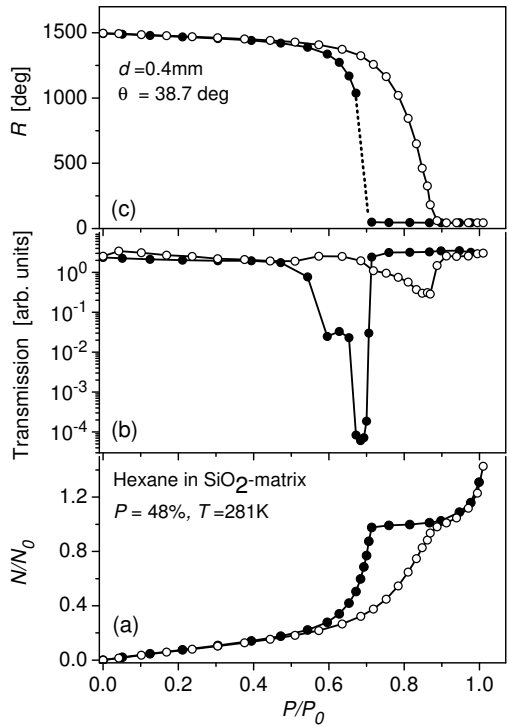, angle=0,width=0.85\columnwidth} \caption{(a) Sorption isotherm of n-hexane condensed in a silica membrane at $T=$281~K along with the simultaneously measured optical transmission (b) and phase retardation (c). Open symbols refer to adsorption, solid symbols to desorption.} \label{fig2}
\end{figure}

For a tilted sample geometry, as used in our experiment, the conversion of the retardation $R$ to the optical birefringence $\Delta n$ reads as:
\begin{eqnarray}
\nonumber
 R&=&\frac{2\pi d(\bar{n}-\Delta
n/3)}{\lambda}\{[1-(\bar{n}+2\Delta n/3)^{-2}\sin^2\theta]^{1/2}
\\
 &-&[1-(\bar{n}-\Delta n/3)^{-2}\sin^2\theta]^{1/2}\}
 \label{eq1}
\end{eqnarray}
where $d$ is the sample thickness, $\bar{n}=(2n_{\rm o}+n_{\rm e})/3$ is the average refractive index, $n_{\rm o}$ and $n_{\rm e}$ are the ordinary and extraordinary refractive indices, respectively. Eq.(\ref{eq1}) is known as the equation for the retardation of the Berek's birefringent compensator\cite{Berek}. Using the literature value of the refractive index of bulk liquid n-hexane ($n_{\rm h}$=1.375) and amorphous SiO$_{\rm 2}$ ($n_{\rm s}$=1.457) \cite{SilicaBulkBir} as well as $n_{\rm v}$=1 for the n-hexane vapor, the magnitude $\bar{n}(f)$ has been evaluated for each filling fraction $f$ by means of the generalized symmetric Bruggeman equation \cite{Bruggeman} for an effective medium. Note that this equation is semi-empirical, it ignores the multipole components of the electric field emerging from the constituents. Eq.(\ref{eq1}) has been solved numerically in order to extract the birefringence magnitude $\Delta n$ for each measured retardation value $R(f)$.

\begin{figure}[htbp]
\epsfig{file=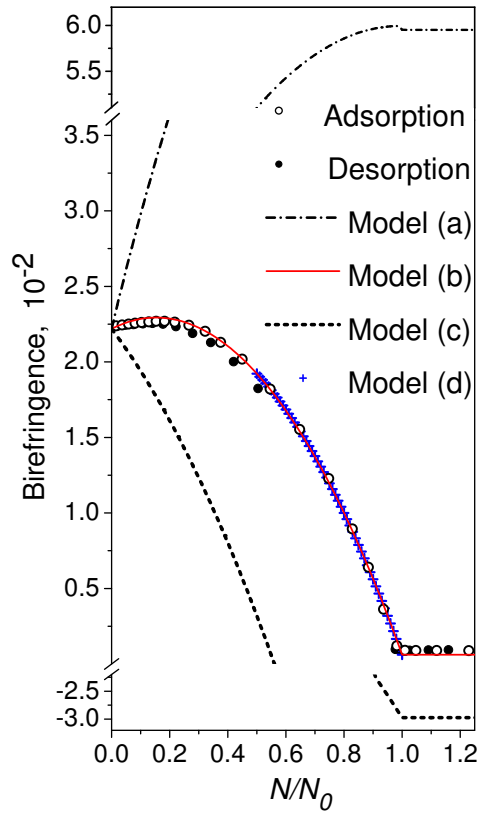, angle=0,
width=0.8\columnwidth}\caption{The optical birefringence $\Delta n$ vs $f=N/N_{\rm 0}$ as determined from the measured phase retardation $R$ (open and solid circles) and calculated by means of the generalized Bruggeman equation for several models of molecular arrangement discussed in the text. The models (a)-(c) correspond to the molecular arrangement as presented in Fig.~\ref{fig5}(a)-\ref{fig5}(c), respectively. The model (d) refers to the capillary condensation regime [see Fig.~\ref{fig4}b] assuming a totally disordered (isotropic) molecular arrangement.} \label{fig3}
\end{figure}

\section{Results and discussion}

The experimental results are shown in Fig.~\ref{fig2}. In parallel with the volumetric sorption data not only the retardation but also the optical transmission of the sample has been recorded.

The volumetric sorption isotherm has the shape well known for many examples of van der Waals molecules in combination with mesoporous substrates. The initial reversible part is due to adsorption of an n-hexane film on the pore walls.  The hysteretic part is typical of capillary condensation and evaporation. The subsequent plateau indicates the completion of pore filling, whereas the final increase is due to the condensation of vapor on the machined rough metal walls of the sample cell and eventually due to bulk condensation. According to theoretical models \cite{Ev,Saam} for cylindrical pores of uniform cross section the adsorption branch should be vertical in the regime of capillary condensation. Both branches have a finite slope. This is an evidence for a broad distribution of pore diameters. Because of the rather gradual increase of the adsorption branch, the filling fraction $f_{\rm c}$ at which capillary condensation starts cannot be determined exactly. We will just refer to an approximate value of 0.5.

The largely different transmission of the adsorption and the desorption branch [Fig.~\ref{fig2}b] in the capillary condensation regime ($f_{\rm c} \le f \le 1)$ are in qualitative agreement with the influence of pore size inhomogeneities and surface roughness within the independent linear channels on the capillary condensation process \cite{Wallacher2004,KaergerpSi}. For adsorption the filling consists of many small pieces of condensate in the narrowest sections of the channels, whereas on desorption - because of pore blocking - some of the condensate resides in pores with larger diameters with terminating menisci in narrow sections. Therefore the average diameter of the filled sections of the pore network is larger than for adsorption. The filled regions are bigger in size and fewer in number whereas the average distance between them is larger and can approach the wavelength of light. This causes strong scattering and hence low transmission along the desorption branch of the hysteresis loop. In fact it was impossible to measure the retardation in this range. Quite similar behavior of the optical transmission has been observed upon n-hexane \cite{Page} and argon \cite{Sopr} condensation in Vycor.

\begin{figure}[htbp]
 \epsfig{file=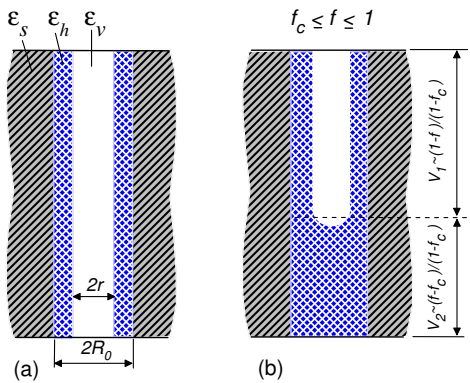, angle=0,
width=0.8\columnwidth}\caption{(a) Schematic sketch of the geometric arrangement of the liquid in a cylindrical nanochannel in the initial, reversible, film-growth regime of the sorption isotherm ($0 \le f \le f_{\rm c}$) and in the hysteretic regime of capillary condensation ($f_{\rm c} \le f \le 1$) (b).} \label{fig4}
\end{figure}

\begin{figure}[htbp]
 \epsfig{file=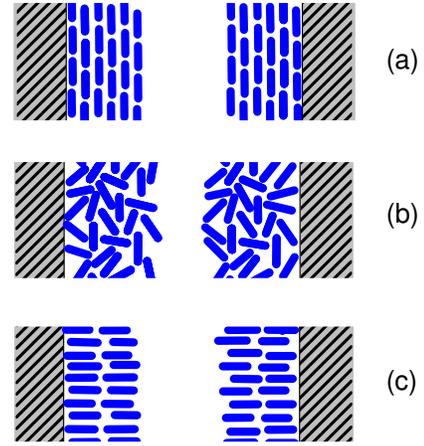, angle=0,
width=0.8\columnwidth}\caption{Three selected models of the molecular arrangements inside the nanochannel as supplement to Fig.~\ref{fig3}. In the panels (a) and (c) fully anisotropic films with molecules aligned parallel to the pore axis (planar orientation) or perpendicular to the pore wall (homeotropic orientation), resp., are schematically sketched.  In panel (b) a totally orientational disordered (isotropic) film is depicted.} \label{fig5}
\end{figure}

The phase retardation $R$ decreases with increase of the fractional filling $f$, see Fig.~\ref{fig2}c, and exhibits the most considerable changes in the capillary condensation regime. Taking into account that the phase retardation is roughly proportional to the optical birefringence $\Delta n$, such behavior can be explained qualitatively considering the fact that the effective average refractive index of the pore filling $\bar{n}_{\rm p}(f)$ gradually approaches the refractive index of amorphous SiO$_{\rm 2}$ $n_{\rm s}$ upon rising $f$. This drops the ''contrast'' of spatial inhomogeneities in the refractive index and thus reduces the birefringence. A more accurate characterization is based on the analysis of the optical birefringence $\Delta n$ which can be directly evaluated from the measured retardation $R(f)$ by means of Eq.(\ref{eq1}). Open and closed circles in Fig.~\ref{fig3} shows $\Delta n$ vs $f$.

In the following we compare the experimental data with an effective medium model. The silica matrix (volume fraction $1-P$) is assumed to contain cylindrical inclusions (volume fraction $P$) that represent the empty, partially or completely filled channels. The cylinder with radius $R_{\rm 0}$ has a vapor filled core with radius $r$ and cylindrical shell of thickness $R_{\rm 0}-r$ filled with liquid n-hexane, see Fig.~\ref{fig4}a. $f$ and $r$ are related via the geometric relation $f=1-r^2/R_{\rm 0}^2$. The permittivity of such an individual inclusion with the field parallel to the pore axis, $\varepsilon_{\parallel}$, is then trivially obtained having the permittivities of vapor and liquid in parallel whereas for $\varepsilon_{\perp}$ we refer to the symmetric Bruggeman equation \cite{Bruggeman, Szultz}:

\begin{eqnarray}
\nonumber
 \varepsilon_{\parallel}&=&(1-f)\varepsilon_{\rm v}+f\varepsilon_{\rm h,\parallel},
\\
 \varepsilon_{\perp}&=&\varepsilon_{\rm h,\perp}\frac{\varepsilon_{\rm v}+\varepsilon_{\rm h,\perp}+
 (1-f)(\varepsilon_{\rm v}-\varepsilon_{\rm h,\perp})}{\varepsilon_{\rm v}+\varepsilon_{\rm h,\perp}-(1-f)(\varepsilon_{\rm v}-\varepsilon_{\rm h,\perp})},
 \label{eq2}
\end{eqnarray}
where $\varepsilon_{\rm v} =n_{\rm v}^2=1$ is the dielectric permittivity of n-hexane vapor, $\varepsilon_{\rm h,\parallel}$ and
$\varepsilon_{\rm h,\perp}$ represent the components of the dielectric permittivity of n-hexane film parallel and perpendicular with respect to the channel axis, respectively. The bulk liquid is of course an optically isotropic medium, due to random orientations of the molecule, but the pore walls may induce some preferred orientation.
For the n-hexane layer three cases, illustrated in Fig.~\ref{fig5}, are investigated, in particular, the isotropic case:
$\varepsilon_{\rm h,\parallel}=\varepsilon_{\rm h,\perp}=n_{\rm h}^2$ (see Fig.~\ref{fig5}b) and the two fully developed anisotropic cases: $\varepsilon_{\rm h,\parallel}=n_{\rm h,e}^2$ and $\varepsilon_{\rm h,\perp}=n_{\rm h,o}^2$ (see Fig.~\ref{fig5}a) or
$\varepsilon_{\rm h,\parallel}=n_{\rm h,o}^2$ and $\varepsilon_{\rm h,\perp}=(n_{\rm h,o}^2+n_{\rm h,e}^2)/2$ (see
Fig.~\ref{fig5}c). The optical anisotropy of a hypothetic n-hexane liquid with complete orientational order can be estimated from the results of calculations of the polarisibility tensor of the n-hexane molecule. One arrives at $n_{\rm h,o}$=1.335 and $n_{\rm h,e}$=1.459 \cite{Knorr}. The effective permittivity $\varepsilon_j^{\rm ef}$ ($j \equiv$ $ \parallel$ or $\perp$) of the composite consisting of silica, liquid n-hexane and n-hexane vapor, is then obtained by solving the generalized Bruggeman equation \cite{Lag,Szultz}:
\begin{eqnarray}
\nonumber
 P\sum_{\rm i=1}^{\rm N}v_{\rm i}\frac{\varepsilon_{\rm i,j}-\varepsilon_{\rm j}^{\rm ef}}{\varepsilon_{\rm j}^{\rm ef}+L_{\rm i,j}(\varepsilon_{\rm i,j}-\varepsilon_{\rm j}^{\rm ef})}
 =(P-1)\frac{3(\varepsilon_{\rm s}-\varepsilon_{\rm j}^{\rm ef})}{2\varepsilon_{\rm j}^{\rm ef}+\varepsilon_{\rm s}}\\
 \label{eq3}
\end{eqnarray}

For the moment all channels are still considered equivalent, as shown in Fig.~\ref{fig4}a. $\varepsilon_{\rm i,j}$ is the pore permittivity as defined in Eq.(2), and $L_{\rm i,j}$ is the depolarization factor of the pore, $L_{\rm i,\parallel}$=0, $L_{\rm i,\perp}$=0.5. $v_i$ refers to the volume fraction of a certain type of partially or completely filled pores, $\sum_{\rm i=1}^{\rm N} v_{\rm i}=1$. If the pores are uniformly filled like in the regime $f<f_{\rm c}$ (see Fig.4a) then one deals with $N=1$ and $v_1=1$. By contrast, in Fig. 4b ($f>f_{\rm c}$) completely and partially filled parts of pores are considered as two independent sets of pores with the volume fractions $v_{\rm 1}$ and $v_{\rm 2}$, respectively. In this case $N=2$ and the magnitudes of the two volume fractions are related to the filling fraction of the pores as stated in the labels of Fig.~4b. The birefringence $\Delta n$ is given by $\sqrt{\varepsilon_{\parallel}^{\rm ef}}-\sqrt{\varepsilon_{\perp}^{\rm ef}}$.
In Fig.~\ref{fig3} we compare the results of the effective medium model for the three cases shown in Fig.~\ref{fig5} with the experimental data. The agreement of the isotropic model (b) with the experiment is almost perfect. The results for the two anisotropic situations are way off, suggesting that the experiment could easily detect even minor deviations from the isotropic case, i.e. from orientational disorder.

Thus, even for the film-adsorbed molecules we found no hints for any collective orientational order. This result is somewhat surprising. For related n-alkane/sapphire (Al$_{\rm2}$O$_{\rm 3}$) interfaces a parallel alignment of the long axes of n-alkane molecules with respect to the substrate surface has been reported \cite{Yeganeh2002}. Also studies on thin films of n-alkanes on planar silica surfaces indicate surface induced alignment effects \cite{Volkmann2002}. In those studies, the surfaces were, however, atomically flat. As mentioned above, we have to assume that the channel walls in our experiments have mean square roughness on the order of the molecular extension. This magnitude seems to be sufficient to smear out any hints of a preferred orientational order in the proximity of the channel walls.\\
This conclusion is supported by a former study that highlights and thoroughly discusses the importance of disorder introduced by the pore wall morphology on the pore-condensed state in matrices of similar type \cite{KaergerpSi}.

So far all pores have been assumed identical with a liquid film of thickness $R_{\rm0}-r$ on the pore walls. According to the current understanding of pore condensation, this geometry should apply to range $0<f<f_{\rm c}$, only, before the onset of capillary condensation. Setting $r=0$, this geometry covers the case of complete filling, $f=1$, too. Perhaps somewhat surprisingly, model (b) also describes well the intermediate regime of capillary condensation, $f_{\rm c}<f<1$,
where completely filled pore segments coexist with other segments that have a liquid film of maximum thickness on the walls, but are otherwise vapor filled, Fig.~\ref{fig4}b. The volume fractions of these two sets of inequivalent pores depend on $f$ as indicated in the figure. The results of the pertinent model (d) are included in Fig.~\ref{fig3}, they differ little from model (b). One concludes that the model employed, though being very sensitive to anisotropies, both shape and intrinsic ones, does not depend much on the spatially arrangement of the components, as long as the volume fractions are considered correctly.  Measurements of the permittivity or of the refractive index may nevertheless give information on the spatial arrangement of the pore filling in case of larger contrast between the components. An analysis of such data had to be based on more refined effective medium models that consider multipolar stray fields.

\section{Conclusions}
In conclusion we presented here the condensation of n-hexane in silica nanochannels as studied by a combination of volumetric sorption isotherms and optical polarimetry measurements. The experimental results are analyzed within the generalized Bruggeman effective medium equation to distinguish the molecular arrangement in different parts of the sorption isotherm. Following  this approach the optical birefringence is found to be independent on a geometrical (positional) arrangement of adsorbed material inside the pores, but is extremely sensitive to the orientational molecular ordering. Thus any kind of intrinsic anisotropy could have been easily detected in principle. However, the results obtained suggest that n-hexane molecules adsorbed in SiO$_{\rm 2}$ nanochannels prepared by thermal oxidation of mesoporous silicon are totally disordered in all regimes of the sorption isotherm, in particular also in the film growth regime.

In particular, the evidence of an entirely isotropic channel confined liquid state, presented here, implicates that the texture found in the crystalline state of n-hexane, discussed in the introduction, is solely dictated by a nano-scale Bridgman mechanism and does not originate in any pre-alignment effects in the nanoconfined liquid.

For the future, it would be important to be able to quantify the wall roughness of the channels in more detail and subsequently to scrutinize its influence on the orientational order in the wall proximity. By the same token, the measurements presented here would profit from an extension of the existing theoretical studies of liquids spatially confined by atomically smooth walls \cite{LitIntroTheo} towards rough walls.

Finally, we envision that the substantial changes in the optical properties of the silica nanochannel array upon controlled filling via the vapor phase, demonstrated here, along with the relative simple way of membrane preparation may stimulate  optical switching applications triggered by capillary condensation/evaporation, similarly as it has been demonstrated in porous photonic superlattices \cite{NPhot}.

\end{document}